\documentclass[sn-mathphys-num]{sn-jnl}

\usepackage{graphicx}%
\usepackage{multirow}%
\usepackage{amsmath,amssymb,amsfonts}%
\usepackage{amsthm}%
\usepackage[title]{appendix}%
\usepackage{xcolor}%
\usepackage{textcomp}%
\usepackage{manyfoot}%
\usepackage{booktabs}%
\usepackage{algorithm}%
\usepackage{algorithmicx}%
\usepackage{algpseudocode}%
\usepackage{listings}%

\usepackage[T1]{fontenc}
\usepackage{bm}
\usepackage{float}
\usepackage{subfigure}
\usepackage{graphicx}
\usepackage{caption}

\usepackage{lmodern}

\begin{document}

\title[Article Title]{Self-Testing Positive Operator-Valued Measurements and Certifying Randomness}


\author[1,2,3]{\fnm{Wenjie} \sur{Wang}}

\author[1,2,3]{\fnm{Mengyan} \sur{Li}}

\author*[1,2,3]{\fnm{Fenzhuo} \sur{Guo}}\email{gfenzhuo@bupt.edu.cn}

\author[4]{\fnm{Yukun} \sur{Wang}}

\author[3]{\fnm{Fei} \sur{Gao}}

\affil[1]{\orgdiv{School of Science}, \orgname{Beijing University of Posts and Telecommunications}, \orgaddress{\city{Beijing}, \postcode{100876}, \country{China}}}

\affil[2]{\orgdiv{Key Laboratory of Mathematics and Information Networks}, \orgname{Beijing University of Posts and Telecommunications, Ministry of Education}, \orgaddress{ \city{Beijing}, \postcode{100876}, \country{China}}}

\affil[3]{\orgdiv{State Key Laboratory of Networking and Switching Technology}, \orgname{Beijing University of Posts and Telecommunications}, \orgaddress{\city{Beijing}, \postcode{100876}, \country{China}}}

\affil[4]{\orgdiv{Beijing Key Laboratory of Petroleum Data Mining}, \orgname{China University of
Petroleum-Beijing}, \orgaddress{\city{Beijing}, \postcode{102249}, \country{China}}}


\abstract{In the device-independent scenario, positive operator-valued measurements (POVMs) can certify more randomness than projective measurements. This paper self-tests a three-outcome extremal qubit POVM in the X-Z plane of the Bloch sphere by achieving the maximal quantum violation of a newly constructed Bell expression $\mathcal{C}_3^{'} $, adapted from the chained inequality $\mathcal{C}_3$. Using this POVM, approximately 1.58 bits of local randomness can be certified, which is the maximum amount of local randomness achievable by an extremal qubit POVM in this plane. Further modifications of $\mathcal{C}_3^{'} $ produce $\mathcal{C}_3^{''} $, enabling the self-testing of another three-outcome extremal qubit POVM. Together, these POVMs certify about 2.27 bits of global randomness. Both local and global randomness surpass the limitations certified from projective measurements. Additionally, the Navascués-Pironio-Acín hierarchy is employed to compare the lower bounds on global randomness certified by $\mathcal{C}_3$ and several other inequalities. As the extent of violation increases, $\mathcal{C}_3$ demonstrates superior performance in randomness certification.}




\maketitle

\section{Introduction}\label{sec1}

Randomness is a vital resource in science and technology, serving as the foundation for a wide range of applications, including numerical simulations \cite{karp1991introduction,li2024smoothing}, machine learning \cite{song2024quantum,li2025efficient}, and cryptographic protocols that ensure data security and privacy in modern communication systems \cite{pirandola2020advances,qin2024decoy}. The generation of truly unpredictable and secure random numbers is of paramount importance, as the quality of randomness directly impacts the reliability and security of these applications. Device-independent (DI) randomness certification offers a robust framework for generating secure random numbers without relying on detailed assumptions about the internal workings of the devices used. Instead, it harnesses the observed correlations between measurement outcomes, such as violations of Bell inequalities, to certify the presence of genuine randomness \cite{acin2007device,gallego2013device,bouda2014device,liu2018device,liu2024quantifying,liu2021device,xiao2023device,mahato2022device,brown2024device,acin2012randomness,woodhead2018randomness,wooltorton2022tight,curchod2017unbounded,nieto2014using,yuan2019quantum,meng2024maximal,borkala2022device,acin2016optimal,andersson2018device}. This approach provides a higher level of security by eliminating potential vulnerabilities arising from device imperfections, misconfigurations, or adversarial manipulation.

Self-testing, which can prove the uniqueness of the quantum state and the required measurements up to a local isometry, is an effective approach to ensure that the obtained randomness is DI. One approach of the self-testing is based on the maximal quantum violation of Bell inequalities \cite{bamps2015sum,kaniewski2016analytic,panwar2023elegant,baccari2020scalable}. The chained Bell inequality $\mathcal{C}_n$ ($n$ represents the number of inputs) was introduced in Ref. \cite{pearle1970hidden} and there has already been a self-testing protocol that proves the quantum state and measurements leading to the maximal violation of the chained Bell inequality are unique up to a local isometry \cite{vsupic2016self}. This suggests that the DI randomness certification based on the maximal quantum violation of $\mathcal{C}_n$ is possible. In Ref. \cite{xiao2023device}, Xiao et al. presented the lower bounds on the DI randomness with respect to the Werner states based on the chained Bell inequality. Their results indicate that 2 bits of global randomness can be certified from the maximal violation of $\mathcal{C}_3$ through projective measurements, showcasing an advantage over other chained Bell inequalities $\mathcal{C}_n(n\neq3)$.

Once the states and measurements are constrained to qudit systems through self-testing, it allows certifying up to $\log_{2}{d}$ bits of local randomness and $2\log_{2}{d}$ bits of global randomness in principle by using projective measurements \cite{dhara2013maximal}. For example, when $d=2$, i.e., when applied to a qubit system, at most 1 bit of local randomness and 2 bits of global randomness can be certified by projective measurements. In addition to projective measurements, positive operator-valued measurements (POVMs) can also be performed on quantum systems. POVMs have the potential to generate more randomness, as they can produce outcomes that exceed the dimensions of the quantum systems they act upon \cite{borkala2022device,pan2021oblivious,acin2016optimal,d2005classical,andersson2018device}. However, the difficulty of self-testing extremal POVMs limits its application to DI randomness. In Ref. \cite{andersson2018device}, a four-outcome extremal qubit POVM was self-tested by the maximal quantum violation of Gisin’s elegant Bell inequality (EBI) \cite{myrvold2009quantum}, and 2 bits of local randomness were certified using this POVM. Woodhead et al. also self-tested the POVMs with four outcomes to a limited degree and certified global randomness restricted to between approximately 3.58 and 3.96 bits \cite{woodhead2020maximal}. Nevertheless, the elements of the above four-outcome qubit POVMs contain components in the Y direction on the Bloch sphere. Due to the involvement of the imaginary unit $i$ in the Pauli operator $\sigma_y$, this may introduce additional complexities in measurement and data analysis. The extremal qubit POVM  restricted in the X-Z plane (labelled as POVM$_{xz}$) has at most three outcomes. Moreover, the three-outcome POVM is more robust against imperfections in the experimental setup than the four-outcome POVM \cite{andersson2018device,gomez2018experimental}, which makes the three-outcome POVM more valuable in practical quantum information processing tasks. A three-outcome extremal qubit POVM$_{xz}$ was self-tested through an inequality $\mathcal{B}_3$ in Ref. \cite{pan2021oblivious}, certifying about 1.58 bits of local randomness, where 1.58 bits is the largest amount of local randomness that can be certified by the extremal qubit POVM$_{xz}$. 

In this paper, we focus on self-testing the new three-outcome extremal qubit POVM$_{xz}$ and performing randomness certification. First, we propose a new Bell expression $\mathcal{C}_3^{'} $ by appropriately modifying $\mathcal{C}_3 $ and self-test a three-outcome extremal qubit POVM$_{xz}$. Using this POVM$_{xz}$, we also certify the most about 1.58 bits of local randomness as in Ref. \cite{pan2021oblivious} by the guessing probability of the eavesdropper Eve. A further modification of $\mathcal{C}_3^{'} $ yields the Bell expression $\mathcal{C}_3^{''} $, from which We certify approximately 2.27 bits of global randomness. The above results break through the limitations of projective measurements in randomness certification. Complementarily, we also use the Navascués-Pironio-Acín (NPA) hierarchy based on projective measurements to compare $\mathcal{C}_3 $ with several other inequalities, including the Bell inequality $\mathcal{B}_3$ \cite{pan2021oblivious}, the Clauser-Horne-Shimony-Holt (CHSH) Bell inequality \cite{clauser1969proposed}, and the EBI \cite{myrvold2009quantum}, in terms of inequality violation for certifying randomness. Different inequalities exhibit different relations between the extent of violation and certifiable randomness. We find $\mathcal{C}_3 $ performs better when the extent of violation is higher. Our work provides modest guidance for selecting inequalities in designing randomness certification experiments. Furthermore, the extremal qubit POVMs we have self-tested can also be applied to other quantum information processing tasks.

\section{Self-testing POVM and local randomness certification}\label{sec2}

We are interested in the following adversarial Bell scenario, consisting of two users, Alice and Bob, and an adversary, Eve. Alice and Bob perform local measurements on their quantum subsystems $\mathcal{H_A}$ and $\mathcal{H_B}$. They share a quantum state $\rho_{AB}\in \mathcal{H_A}\otimes \mathcal{H_B}$. We denote by $A_i$ $(i=1,...,h)$ and $B_j$ $(j=1,...,m)$ the observables of Alice and Bob and assume that they have outcomes $a \in\{1,...,u\}$ and  $b \in\{1,...,v\}$, respectively. The joint probability that the outputs $a$ and $b$ are obtained given the inputs $A_i$ and $B_j$ is 
\begin{equation}
G(a,b)=p(ab|A_i B_j)=tr(\rho_{AB}\,A_{a|i}\otimes B_{b|j}).
\end{equation}
A total of $h\times m\times u\times v$ such probabilities form a set $\mathcal{P}=\{p(ab|A_i B_j)\}$, which we define as a $\textit{behavior}$.  

The goal of Eve is to guess Alice's outcome $a$ with maximal probability. To achieve this, Eve can prepare Alice’s and Bob’s system in any way compatible with the given behavior $\mathcal{P}$ and the laws of quantum physics. Eve may also try to entangle with the subsystems of Alice and Bob and create correlations which would allow her to gain some information about the outcomes of the experiment. Eve's strategy consists of a POVM $F$ on $\mathcal{H_E}$ with elements $F_a$ whose result is Eve’s best guess on Alice’s outcome. Such a strategy is compatible with $\mathcal{P}$ if
\begin{equation}
p(ab|A_i B_j)=tr(\rho_{ABE}\,A_{a|i}\otimes B_{b|j}\otimes I_E),
\end{equation}
where $\rho_{ABE}\in \mathcal{H_A}\otimes \mathcal{H_B}\otimes \mathcal{H_E}$ and $\rho_{AB}=tr_E (\rho_{ABE})$. Then the local guessing probability of Eve can be expressed as
\begin{equation}
\begin{aligned}
G(i,\mathcal{P})&=\mathop{max}\limits_{F} \sum\limits_a p(a,a|A_{a|i},F_a)\\
&=\mathop{max}\limits_{F} \sum\limits_a tr(\rho_{ABE}\,A_{a|i}\otimes I_B \otimes F_a).
\end{aligned}
\end{equation}
The randomness can be quantified by the $\textit{min-entropy}$ $H_{\min}=-\log_{2}{G(i,\mathcal{P})}$.

In Ref. \cite{vsupic2016self}, the self-testing protocols of $\mathcal{C}_n$ has been completed. For $n = 3$, the expression is
\begin{equation}
\mathcal{C}_3=\langle A_1 B_1\rangle +\langle A_2 B_1\rangle+\langle A_2 B_2\rangle+\langle A_3 B_2\rangle+\langle A_3 B_3\rangle-\langle A_1 B_3\rangle\leq 4.
\end{equation}
The maximal quantum violation of $\mathcal{C}_3$ is 
\begin{equation}
\mathcal{C}_3^{max}=3\sqrt{3},
\end{equation}
with the maximally entangled two-qubit state 
\begin{equation}
|\phi_+\rangle=\frac{1}{\sqrt{2}}(|00\rangle+|11\rangle),
\end{equation}
and the measurements as follows,
\begin{equation}
\begin{aligned}
A_1&=\sigma_z,\qquad\qquad\qquad \,\,\,B_1=\frac{1}{2}\sigma_x+\frac{\sqrt{3}}{2}\sigma_z,\\
A_2&=\frac{\sqrt{3}}{2}\sigma_x+\frac{1}{2}\sigma_z,\qquad B_2=\sigma_x,\\
A_3&=\frac{\sqrt{3}}{2}\sigma_x-\frac{1}{2}\sigma_z,\qquad B_3=\frac{1}{2}\sigma_x-\frac{\sqrt{3}}{2}\sigma_z,
\end{aligned}
\end{equation}
where $\sigma_x$ and $\sigma_z$ are the Pauli operators. We mark this set of settings as  $\{|\phi_+\rangle, A_i,B_i\}(i=1,2,3)$. If there exists another set of settings $\{|\psi^{'}\rangle, A_i^{'},B_i^{'}\}(i=1,2,3)$ also capable of achieving the maximal quantum violation of $\mathcal{C}_3$, then they are equivalent, i.e., there exists a local isometry 
\begin{equation}
\Phi=\Phi_A\otimes\Phi_B\otimes I_E : \mathcal{H_A}\otimes \mathcal{H_B}\otimes \mathcal{H_E}\rightarrow (\mathcal{H_A}\otimes \mathcal{H}_2)\otimes(\mathcal{H_B}\otimes \mathcal{H}_2)\otimes\mathcal{H_E},
\end{equation}
such that 
\begin{equation}
\begin{aligned}
\label{9}
&\Phi(|\psi^{'}\rangle)=|\chi\rangle\otimes|\phi_+\rangle, \\
&\Phi(A_i^{'}\otimes B_i^{'}|\psi^{'}\rangle)=(A_i\otimes B_i |\phi_+\rangle)|\chi\rangle,
\end{aligned}
\end{equation}
for some $|\chi\rangle\in \mathcal{H_A}\otimes \mathcal{H_B}\otimes \mathcal{H_E}$.

Based on the measurement setups for $\mathcal{C}_3$ mentioned above, we introduce a three-outcome measurement $A_4$ on Alice’s side, which will be used to generate local randomness. Given this configuration of measurements, we define the modified chained Bell expression 
\begin{equation}
\mathcal{C}_3 ^{'}=\mathcal{C}_3-\alpha\left[ p(1,1|A_4,B_1)+p(2,-1|A_4,B_2)+p(3,1|A_4,B_3)\right],
\end{equation}
where $\alpha$ is an arbitrary strictly positive constant. Since the last three terms on the right side of the equation are always negative, both $\mathcal{C}_3 ^{'}$ and $\mathcal{C}_3$ have the same classical upper bound. The same argument implies that the quantum violation of $\mathcal{C}_3 ^{'}$ cannot be larger than $3\sqrt{3}$, i.e., 
\begin{equation}
\mathcal{C}_3^{' max}\leq3\sqrt{3}.
\end{equation}

The only condition under which the above expression holds with equality is that the probabilities $p(1,1|A_4,B_1)$, $p(2,-1|A_4,B_2)$, and $p(3,1|A_4,B_3)$ are all zero. This occurs when the POVM elements ${A_{k|4}}$ (for $k=1,2,3$) are antialigned with the three projective measurements $B_{1|1}$, $B_{-1|2}$, and $B_{1|3}$ on Bob’s side, respectively. In this case, we obtain
\begin{equation}
\begin{aligned} 
&A_{1|4}=\frac{2I-\sigma_x-\sqrt{3}\sigma_z}{6}, \\
&A_{2|4}=\frac{I+\sigma_x}{3},\\
&A_{3|4}=\frac{2I-\sigma_x+\sqrt{3}\sigma_z}{6}.
\end{aligned}
\end{equation}
Thus, the modified chained Bell expression $\mathcal{C}_3 ^{'max} = 3\sqrt{3}$ certifies the POVM $A_4=\{A_{k|4}\}(k=1,2,3)$. Since the elements of $A_4$ are rank-one and linearly independent, according to Ref. \cite{d2005classical}, $A_4$ is a three-outcome extremal qubit POVM.

It is straightforward to calculate that if Alice performs the POVM $A_4$ on her subsystem, each outcome occurs with an equal probability of $\frac{1}{3}$. To examine the presence of Eve, we demonstrate the unpredictability of Alice's outcomes in a DI way. As mentioned earlier, the probability that Eve perfectly guesses Alice's outcomes is given by 
\begin{equation}
G(4,\mathcal{P}) =\mathop{max}\limits_{F} \sum\limits_{k=1}^{3} p(k,k|A_{k|4},F_k).
\end{equation}
The following proof confirms that Eve's guessing probability is $\frac{1}{3}$, thereby validating the unpredictability of Alice's outcomes.

A family of qubit POVM operators $A_{k|4}$ can generally be expressed as
\begin{equation}
\label{14}
A_{k|4} =\gamma_k^0 I+\gamma_k^1\sigma_z+\gamma_k^2\sigma_y+\gamma_k^3\sigma_x,
\end{equation}
where $\sigma_x$, $\sigma_y$, and $\sigma_z$ are the Pauli operators. The coefficients $\gamma_k^t$ with $t= 0, 1, 2, 3$ take the form
\begin{equation}
\begin{aligned} 
\label{15}
&\gamma_k^0=p(k|A_{k|4}), \\
&\gamma_k^1=\frac{1}{\sqrt{3}}(E_{k|4,1}-E_{k|4,3}),\\
&\gamma_k^2=-E_{k|4,1}+E_{k|4,2}-E_{k|4,3},\\
&\gamma_k^3=E_{k|4,2},
\end{aligned}
\end{equation}
where $E_{k|4,j}=\sum\limits_b b\,p(k,b|A_4,B_j)$.

On the support of $\Phi_A$, each element of $A_4$ can be represented by an operator $\tilde{A}_{k|4}$ acting on $\mathcal{H_A}\otimes \mathcal{H}_2$,
\begin{equation}
\tilde{A}_{k|4} =\tilde{A}_{k|4}^0\otimes I+\tilde{A}_{k|4}^1\otimes\sigma_z+\tilde{A}_{k|4}^2\otimes\sigma_y+\tilde{A}_{k|4}^3\otimes\sigma_x,
\end{equation}
where each $\tilde{A}_{k|4}^t$ is a Hermitian operator on $\mathcal{H_A}$. Assume $|\psi\rangle$ is any quantum state shared by Alice, Bob, and Eve that can achieve the maximal quantum violation of $\mathcal{C}_3$. From the self-testing relations given by Eq. (\ref{9}), it follows that
\begin{equation}
\begin{aligned} 
&\Phi(B_1|\psi\rangle)=\frac{1}{2}\{|\chi\rangle\otimes[I\otimes(\sigma_x+\sqrt{3}\sigma_z)|\phi^+\rangle]\}, \\
&\Phi(B_2|\psi\rangle)=|\chi\rangle\otimes(I\otimes\sigma_x|\phi^+\rangle),\\
&\Phi(B_3|\psi\rangle)=\frac{1}{2}\{|\chi\rangle\otimes[I\otimes(\sigma_x-\sqrt{3}\sigma_z)|\phi^+\rangle]\},
\end{aligned}
\end{equation}
and then
\begin{equation}
\begin{aligned}
\label{18}
&\gamma_k^0=\langle\psi|\tilde{A}_{k|4}\otimes I|\psi\rangle=\langle\chi|\tilde{A}_{k|4}^0|\chi\rangle, \\
&\gamma_k^1=\frac{1}{\sqrt{3}}\langle\psi|\tilde{A}_{k|4}\otimes (B_1-B_3)|\psi\rangle=\langle\chi|\tilde{A}_{k|4}^1|\chi\rangle,\\
&\gamma_k^2=\langle\psi|\tilde{A}_{k|4}\otimes (B_1-B_2+B_3)|\psi\rangle=\langle\chi|\tilde{A}_{k|4}^2|\chi\rangle,\\
&\gamma_k^3=\langle\psi|\tilde{A}_{k|4}\otimes B_2|\psi\rangle=\langle\chi|\tilde{A}_{k|4}^3|\chi\rangle.
\end{aligned}
\end{equation}

We now introduce the normalized state
\begin{equation}
|\phi_{A^{'}}^k\rangle=\frac{F_k|\chi\rangle}{\sqrt{q_k}}.
\end{equation}
Without loss of generality, Eve’s measurement $F_k$ is considered to be a projective measurement. In this way, it results in
\begin{equation}
\label{20}
\gamma_k^t=\langle\chi|\tilde{A}_{k|4}^t|\chi\rangle=\sum\limits_{k^{'}}\langle\chi|F_{k^{'}}\tilde{A}_{k|4}^tF_{k^{'}}|\chi\rangle=\sum\limits_{k^{'}}q_{k^{'}}\langle\phi_{A^{'}}^{k^{'}}|\tilde{A}_{k|4}^t|\phi_{A^{'}}^{k^{'}}\rangle=\sum\limits_{k^{'}}q_{k^{'}}\beta_k^{t,k^{'}},
\end{equation}
where we have defined the coefficients $\beta_k^{t,k^{'}}=\langle\phi_{A^{'}}^{k^{'}}|\tilde{A}_{k|4}^t|\phi_{A^{'}}^{k^{'}}\rangle$. Note that these coefficients define a family of qubit POVMs $U^{k^{'}}$ with operators $U_k^{k^{'}}=\sum\limits_t \beta_k^{t,k^{'}}\sigma_t$. From Eq. (\ref{14}) and Eq. (\ref{20}), we can derive that $A_{k|4}=\sum\limits_t\sum\limits_{k^{'}} q_{k^{'}}\beta_k^{t,k^{'}}\sigma_t=\sum\limits_{k^{'}} q_{k^{'}}U_k^{k^{'}}$, which implies the existence of a convex combination of the original POVM $\{A_{k|4}\}$ in terms of the POVMs $\{U_k^{k^{'}}\}$ with respective weights $q_{k^{'}}$. However, since $A_{k|4}$ is extremal, it leads to $\beta_k^{t,k^{'}}=\gamma_k^t$ for all $k^{'}$. Particularly, $\beta_k^{0,k}=\gamma_k^0=\frac{1}{3}$ for all $k$. Accordingly, Eve's guessing probability is given by
\begin{equation}
\begin{aligned}
\label{21}
G(4,\mathcal{P}) &=\sum\limits_{k} p(k,k|A_{k|4},F_k)\\
&=\sum\limits_{k}\langle\psi|\tilde{A}_{k|4} F_k|\psi\rangle\\
&=\sum\limits_{k}\langle\chi|\tilde{A}_{k|4}^0 F_k|\chi\rangle\\
&=\sum\limits_{k}q_k\beta_k^{0,k}\\
&=\frac{1}{3}.
\end{aligned}
\end{equation}
Thus we can certify $-\log_{2}{\frac{1}{3}}\approx1.58$ bits of local randomness.

\section{Self-testing POVM and global randomness certification}\label{sec3}

Motivated by the previous discussion, we investigate the possibility to certify global randomness from the chained Bell inequality using POVMs. For this purpose, it is necessary to add a three-outcome measurement $B_4$ on Bob's side based on $\mathcal{C}_3 ^{'}$ and construct the following modified expression,
\begin{equation}
\begin{aligned}
\mathcal{C}_3 ^{''}&=\,\mathcal{C}_3 ^{'}-\beta\left[ p(-1,1|A_1,B_4)+p(1,2|A_2,B_4)+p(-1,3|A_3,B_4)\right]\\&=\,\mathcal{C}_3-\alpha\left[ p(1,1|A_4,B_1)+p(2,-1|A_4,B_2)+p(3,1|A_4,B_3)\right]\\&\qquad\,\,\,-\beta\left[ p(-1,1|A_1,B_4)+p(1,2|A_2,B_4)+p(-1,3|A_3,B_4)\right],
\end{aligned}
\end{equation}
where $\beta$ is an arbitrary strictly positive constant. Obviously, $\mathcal{C}_3 ^{''}$ also has the same classical upper bound as $\mathcal{C}_3$, and the maximal quantum violation of $\mathcal{C}_3 ^{''}$ cannot exceed that of $\mathcal{C}_3$, i.e.,
\begin{equation}
\mathcal{C}_3^{'' max}\leq3\sqrt{3}.
\end{equation}
The equality holds if and only if all probabilities $p(1,1|A_4,B_1)$, $p(2,-1|A_4,B_2)$, $p(3,1|A_4,B_3)$, $p(-1,1|A_1,B_4)$, $p(1,2|A_2,B_4)$, and $p(-1,3|A_3,B_4)$ are zero. Similarly, we can test the three-outcome
extremal qubit POVM $B_4=\{B_{l|4}\}(l=1,2,3)$ on Bob's side from the antialignment condition as follows:
\begin{equation}
\begin{aligned} 
&B_{1|4}=\frac{I+\sigma_z}{3}, \\
&B_{2|4}=\frac{2I-\sqrt{3}\sigma_x-\sigma_z}{6},\\
&B_{3|4}=\frac{2I+\sqrt{3}\sigma_x-\sigma_z}{6}.
\end{aligned}
\end{equation}

Nevertheless, when performing joint measurements, the coefficients of the joined POVMs cannot be represented as done in Eq. (\ref{18}). We turn to quantify the global randomness resulting from $A_4$ and $B_4$ through the joint probability
\begin{equation} 
G(k,l)=\mathop{max}\limits_{k,l} p(k,l|A_4,B_4).
\end{equation}
We can get $G(k,l)=\frac{1}{9}(1+\frac{\sqrt{3}}{2})$. Quantifying this in bits gives approximately 2.27 bits of randomness.

I believe it is necessary to clarify here that, for other chained Bell inequalities where $n\neq3$, we currently consider it impossible to construct a suitable extremal POVM as described above. Whether there is a better approach remains an open question.

\section{Randomness certification with NPA}\label{sec4}

In this section, we investigate the lower bounds on DI global randomness by imposing constraints of Bell inequality violation on the NPA hierarchy. For any observed violation $V$ of a given Bell inequality, the maximum guessing probability $P^{*}(ab|A_iB_j)$ can be expressed as the solution to the following optimization problem \cite{xiao2023device}:
\begin{equation}
P^{*}(ab|A_iB_j)=\max\limits_{\{\rho, A_{a|i}, B_{b|j}\}}p(ab|A_i B_j),
\end{equation}
\begin{equation}
\mathrm{s.t.} \quad\sum\limits_{abij}c_{abij} P(ab|A_iB_j)=V,
\end{equation}
\begin{equation}
p(ab|A_iB_j)=tr(\rho\,A_{a|i}\otimes B_{b|j}),
\end{equation}
where the optimization is defined over Hilbert spaces of arbitrary dimensions and encompasses all states $\rho$ and all measurement operators $A_{a|i}$ and $B_{b|j}$. 

\begin{figure}[h]
\centering
\includegraphics[width=0.8\textwidth]{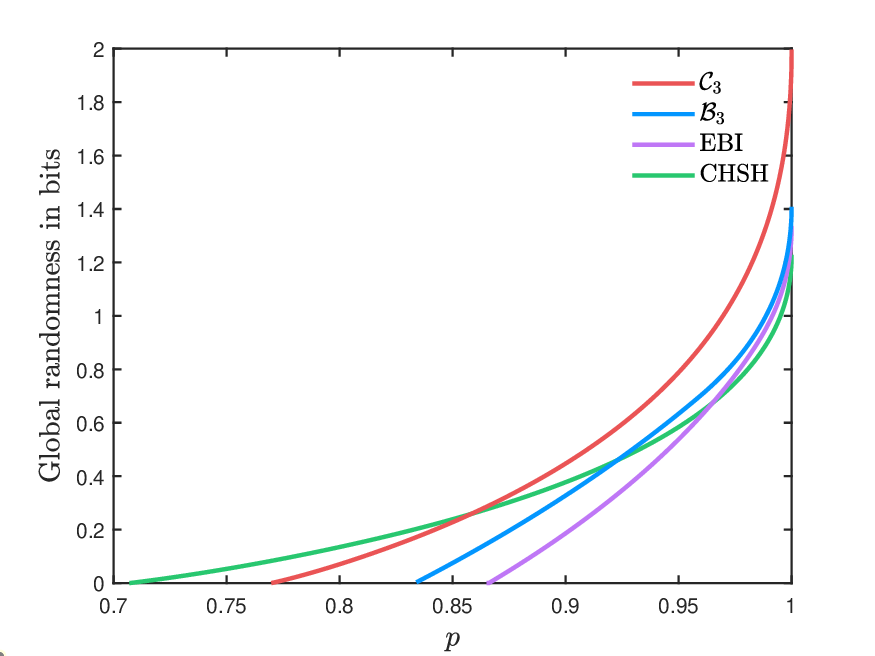}
\caption{Comparison of the lower bounds on DI global randomness between the chained Bell inequality $\mathcal{C}_3$,  the Bell inequality $\mathcal{B}_3$, the CHSH Bell inequality, and the EBI. All the curves were obtained at the Q$_2$ level of the NPA hierarchy.}\label{fig1}
\end{figure}

As shown in Fig. \ref{fig1}, we compare the lower bounds of certifiable randomness for inequalities $\mathcal{C}_3$, $\mathcal{B}_3$, EBI, and CHSH based on inequality violation using the NPA hierarchy. The parameter $p$ quantifies the degree of violation for each inequality. When $p=1$, the maximal quantum violation of each inequality is attained. To provide a more intuitive understanding, consider one specific case with $\mathcal{C}_3$: the correlations take the form of $\mathcal{Q}=p\mathcal{R}+(1-p)\mathcal{S}$, where $\mathcal{Q}$ represents the quantum correlations yielding the maximal value of $3\sqrt{3}p$ from the Werner state, and $\mathcal{R}$ represents the quantum correlations that produce the maximal quantum violation of $3\sqrt{3}$ for the inequality $\mathcal{C}_3$, while $\mathcal{S}$ represents completely random correlations. From the figure, it can be observed that randomness can be extracted in the $\mathcal{C}_3$ scenario only when the inequality $\mathcal{C}_3$ is violated, i.e., when $p>4/(3\sqrt{3})\approx 0.7698$ \cite{xiao2023device}. The same applies to the other three inequalities $\mathcal{B}_3$, EBI, and CHSH.

Moreover, Fig. \ref{fig1} illustrates that for $p>0.86$, the red curve ($\mathcal{C}_3$) is positioned above the other curves (the three other inequalities), demonstrating that $\mathcal{C}_3$ certifies more randomness when the degree of violation is higher. Furthermore, when $p=1$, corresponding to points of maximal quantum violation, the chained inequality $\mathcal{C}_3$ achieves the theoretical optimal value of 2 bits. This finding aligns with the results reported in Ref. \cite{xiao2023device}. We can observe that not all inequalities can achieve 2 bits of randomness, suggesting that $\mathcal{C}_3$ has an advantage in certifying randomness according to the NPA.

\section{Conclusion}\label{sec13}

In this paper, we worked on self-testing the three-outcome extremal qubit POVM in the X-Z plane and investigated its application in randomness certification. By modifying the chained inequality $\mathcal{C}_3$, we proposed a new Bell expression $\mathcal{C}_3^{'}$. At the point of maximal quantum violation of $\mathcal{C}_3^{'}$, a three-outcome extremal qubit POVM$_{xz}$ was self-tested. By incorporating Eve into the Bell test scenario and leveraging the self-testing properties of $\mathcal{C}_3$, we certified approximately 1.58 bits of local randomness, based on Eve's guessing probability. This is the maximum value of local randomness that can be certified by an  extremal qubit POVM$_{xz}$. Subsequently, we further modified $\mathcal{C}_3^{'}$ to obtain $\mathcal{C}_3^{''}$. Additionally, we derived a three-outcome extremal qubit POVM$_{xz}$ for Bob, which, together with the three-outcome extremal qubit POVM$_{xz}$ for Alice, certified approximately 2.27 bits of global randomness. The above results surpass the limitations of projective measurements in randomness certification, demonstrating that POVMs offer advantages in this area.
The Bloch vectors of the extremal POVMs we have self-tested equally spaced in the X-Z plane with robustness in the experiments. In addition, the extremal POVMs can also be applied to other aspects of quantum information science. Furthermore, leveraging the NPA hierarchy, we compared the lower bounds on global randomness certification for inequalities $\mathcal{C}_3$, $\mathcal{B}_3$, EBI, and CHSH based on the extent of violation. We found that the relationship between the amount of randomness certifiable by different inequalities and the extent of the violation varies, with $\mathcal{C}_3$ certifying more randomness as the extent of the violation increases. This provides some guidance for designing randomness certification experiments to select inequalities.

\begin{itemize}
\item This article has been submitted to "Quantum Information Processing" on 22 November 2024.
\end{itemize}

\bibliography{sn-bibliography}

\end{document}